# Convivial Conversational Agents – shifting toward relationships

*Rafael A. Calvo[1], Dorian Peters[1,2]*
1 Dyson School of Design Engineering, Imperial college London
2 Institute for Technology and Humanity, University of Cambridge

*Designing conversational agents to take into account their influence on relationships, autonomy, and community, will improve both their efficacy and impact.*

**Key Insights**
- The need for health services is increasing faster than health budgets, so companies and governments are seeking solutions that leverage conversational agents and generative AI.
- In the pursuit of scalable agentic solutions, we may alienate human support networks and dehumanise those we are aiming to help. The concept of 'conviviality' helps address these concerns.
- We need to design convivial conversational agents which requires empirical participatory approaches that centre human autonomy, relationships and community.

## Introduction

conversational AI (CAI) systems offer opportunities to scale service provision to unprecedented levels and governments and corporations are already beginning to deploy them across services. The economic argument is similar across domains: use CAI to automate the time-consuming conversations required for customer, client or patient support. Herein we draw on our work in dementia care to explore some of the challenges and opportunities for CAI, and how a new way of conceptualising these systems could help ensure essential aspects for human thriving are not lost in the process of automation.

Providing personalised support to people living with dementia (PLWD) and their families is both difficult and expensive. The UK has nearly 1 million PLWD while the US will soon have 1 million new cases every year (Fang, 2025). These individuals and their families would generally rather live at home for as long as possible, but this requires a level of home care that is not currently available--or even feasible--at scale. To reduce the burden on specialist carers and increase access to homecare, our team is among those experimenting with ways conversational agents could be developed to provide some aspects of this support. For example, CAI can provide tech support for smart home care systems [27], automate regular checkups, react with instant first-line support to urgent needs, or offer interventions to support family wellbeing [24].

While CAI offers these opportunities, it also raises many questions, including around safety and privacy (which have been explored in depth elsewhere [e.g. 29]). In addition, we wish to posit a new series of questions: those that arise from the ways *relationships and interactions between humans* are affected by conversations with software agents, and which can only be fully understood at a systems level. While technology can generally be scaled with more servers and data, it is important to ask: what is the impact of such scaling on the richness and quality of human relationships? What are the negative consequences, and can they be prevented?

We focus herein on the challenges that CAI introduces to community healthcare, but many of these challenges arise across other domains. This is because conversational agents are social agents-- in fact, progress made in voice generation and linguistic fluency makes them increasingly indistinguishable from humans. They are becoming social companions and advisors, including to patients, carers, and doctors [7, 26]. Just as a new care provider affects the relationships between existing ones, these agents will do the same. For example, they may provide information that contradicts medical or family advice or that misaligns with the values of the community in which a patient lives. They may be too forceful, insensitive or otherwise inappropriate in ways that disengage patients and carers from activities important for their health. They could enhance, but may equally diminish, existing human-to-human health conversations by displacing or inadvertently discouraging them. They could reduce the quality of communication between patient and health worker or between a patient and their family or community. Without a systems view of these social dynamics, and a relationship-centred lens, we will fail to anticipate or even identify social ecosystem harms when they arise.

Therefore, we bring two different disciplinary perspectives to this new design challenge. First, we point to the critique of automation using the concept of 'conviviality' introduced by Ivan Illich [9] decades ago, in order to raise the issue of how automation can affect our capacity to live together harmoniously, autonomously and interdependently. This critique can be used to rethink how we might design cutting-edge "convivial tools", including convivial conversational agents, in ways that preserve both individual autonomy and relatedness.

The second lens we apply is from social psychology, and provides empirical and theoretical evidence of how autonomy and relatedness are fundamental to psychological health and how they can be supported through design [18, 19]. We discuss this through concrete examples of tools to help families and community health workers supporting people living with dementia.

## The concept of conviviality – living together

> *Only within limits can machines take the place of slaves; beyond these limits they lead to a new kind of serfdom… Once these limits are recognised, it becomes possible to articulate the triadic relationship between persons, tools, and a new collectivity. Such a society, in which modern technologies serve politically interrelated individuals rather than managers, I will call 'Convivial'…" I have chosen 'convivial' as a technical term to designate a modern society of responsibly limited tools"* – Ivan Illich *[9]*

In the text from which this excerpt is drawn, Illich introduces his critique of industrialisation. In particular, he discusses the impact of industrial tools and services on three factors that we consider especially relevant to the design of CAI and paraphrase below:
1) **Tolerant coexistence**, or what in Spanish is called "Convivencia": our ability to live together effectively,
2) **Autonomy** – the flip side to "serfdom" which preserves individual self-determination.
3) **Relatedness** – the reality and essential importance of relationships and interdependence

Illich's concerns, raised back in 1973 at the time of both Weizenbaum's Eliza chatbot and the Jetsons TV Show, reflected the tensions of a society negotiating whether technologies were subordinate tools or controlling masters. Since then, Illich's ideas have been adopted and applied by social scientists [8], education researchers [6] and architects, among others. This tension is perhaps even more salient in our own times. We still see a spectrum of opinions about the impact of AI which spans from radical optimism ("AI will fix everything") to existential dread ("AI will enslave us"). Coming in closer to the latter, Illich [9] described "hyperindustrial" or "overefficient" tools growing out of control and becoming master and executioner. These ideas on technologically driven serfdom were very controversial at the time [14] but have actually become less so today as our lives become increasingly shaped by digital technologies. As we stand in the shadow of, not only dystopian superintelligence narratives, but more importantly, very real and growing power disparities, Illich's concerns grow ever more plausible. As such, it makes sense to reexamine his vision for technology to see what we might productively appropriate toward creating more human-centred, relationship-centred and "convivial" AI.

Illich's use of the term 'conviviality' derives from "convivencia", a Spanish word referring to the ability to live together harmoniously, such as in a household or community. It also encompasses the broader concept of tolerant coexistence (e.g. historically, the "Convivencia" in Spain describes a period of inter-religious coexistence between Muslims, Jews and Christians, that spanned the 8$^{th}$ to the 15$^{th}$ centuries). As such, Alberto Escobar [5] draws on conviviality as a component of what he proposes as 'pluriversal design' -- a philosophy and politics that enable "a world of many worlds" which emphasises ontological and epistemological diversity.

In essence, conviviality acknowledges that, in order for a group to flourish, we need to consider relations among members and the outside world, as well as the autonomy of individuals within the group. In this way relatedness and autonomy are intertwined, a conceptualisation independently reflected in the psychological work of Ryan and Deci [21] (which we come to later). Illich's primary concerns for relatedness and human autonomy both fall directly within the sphere of influence of AI. While individual autonomy has received some focus in HCI [1], relationality demands more attention. CAI's language capabilities allow it to intervene in human-like ways and potentially cause effects that only emerge at a group level. These may not be visible at the individual level of analysis most common in human-computer interaction. In other words, while an individual might benefit from using a tool at an immediate practical level, a relationship or a group might suffer; for example, if carers disengage, patients become more isolated, or human support networks dissolve. Recently, within HCI, Karusala and Anderson use conviviality as a lens to understand social media use for health, presenting examples that both support and hinder users' ability to help their families and communities.[11]

Voinea [28] has argued that Illich follows a systemic approach by combining the analysis of the individual and the group simultaneously. For example, within educational research, Hollingworth [8] studied organisational and first-person phenomenological aspects of what makes a school convivial. Through interviews and ethnographic work Hollingworth explores the conditions that bridge social and ethnic divide, including individual conditions (e.g. academic interests) and relational conditions (e.g. friendships).

In contrast, within modern digital experience, it is often *non-convivial* interactions, like filter bubbles [17] and misinformation, which dominate and likely contribute to social division and anti-cosmopolitan trends. Governments are increasingly concerned about the role of AI in exacerbating such social problems which undermine harmonious coexistence. These concerns are no less significant for health-service provision because these services are essential to community flourishing.

Cipolla [3] posits that interpersonal relations are an essential component of such 'relational services' (those for which people interact to produce a commonly recognised benefit). Extrapolating from Cipolla's case studies, we could consider 'home-based dementia nursing' a relational service. The nurse offers care to a patient and could be considered the service provider. But the service being performed is by no means a faceless transaction. Instead, by taking place within the person's home and over an extended period, the interaction generates a personal relationship. The nurse and patient may share meals, personal stories and intimate moments--the service changes both lives. When a conversational agent is introduced into this relationship, what changes?

Conversational agents already affect 'household politics' by suggesting behaviours, providing information, and connecting to other 'agents' (human or otherwise). These interventions are shaped by external values determined by the designers, datasets and/or profit incentives that shape the technology. These values may be at odds with those embodied by the relationships within which the CAI intervenes. In this way, CAI can affect relationships of power and interdependence, for example between carers and the cared for, young and old, parents and children, husbands and wives--by framing, presenting or recommending courses of action [25].

Individuals and communities around the world are understandably sensitive about values relating to health, for example, regarding end-of-life choices, reproductive rights, caretaking, and other highly value-laden topics. Value sensitive design can be employed to design CAI in participatory ways [e.g. 23]. However, the concept of conviviality does not simply equate to values alignment. In fact, it makes values-based assumptions: that coexistence, human autonomy and interrelatedness should be supported by our tools. These ideas have shaped innovations in education and architecture but have yet to be leveraged for more relational CAI. Of course, one might reasonably suggest that the best way to preserve human relationships may be to avoid the introduction of AI altogether. No doubt, this will be the best option in certain contexts. However, there is also reason to believe that the careful application of CAI for healthcare may be our best path to improving an increasingly dire status quo, making a robust study of its effective use a moral imperative.

## The moral imperative for scaling service provision

In every measure, and across the globe, the demand for health services far outstrips available supply. The "Tilt towards technology", particularly AI [e.g. 4], is therefore, a logical pathway to explore for helping address these challenges.

In the context of dementia care, the problem is growing as lifespans do. In the UK almost 1 million people live with dementia and face a massive shortfall of health workers. For example, an intervention such as a care ecosystem [13] requires up to 1 case officer for every 40 patients. To meet demand, the health system would require 25,000 case officers to care for all PLWD in the country (and this is in addition to the specialist physicians, nurses and other health workers also needed). Mobilising such numbers is not economically feasible, and even if it were, it's not clear there would be enough people available to take on these roles. Furthermore, this sobering reality reflects a best-case scenario: healthcare in a high-income country. For low- and middle-income countries, in which health systems are under even more strain, the problem is significantly greater. It's therefore imperative that care systems and technologists look for ways technology might help to meet some of this need.

However, the design of relational services must be wary of dehumanised measures of success. Services, such as community healthcare, are often measured by the number of "engagements"-- points of contact or conversations--between carers and patients or their families. For example, in the UK, community healthcare organisations engage in over 100 million 'care contacts' per year.[15] They can also be measured by the number of unplanned visits, or hospitalisations. A narrow focus on "scaling-up" to increase "engagements" without attention to the quality or efficacy of these engagements, or the system-level side-effects, could lead to wasting resources on technologies that widen, rather than narrow, the gap in services.

Returning to our dementia care example, we might view 'home-based dementia nursing' as a convivial relational environment. In doing so, our attention is drawn beyond nursing as a *transaction* to nursing as a nourishing and supportive *relationship*. Therefore 'Optimising' within this context demands maintaining or improving the quality and impact of this relationship, rather than on a count of transactions. While the term conviviality has been used in multi-agent system research, we are not aware of any measures of conviviality in Illich's terms, so we discuss opportunities for using existing theoretically-aligned measures below.

## Grounding Convivial AI – Theory and Methods

How might we begin to research, develop and evaluate conversational AI through a convivial lens? Although Illich's work did not go so far as to propose methods or measures for convivial tools, there are aligned theories with their own methods that could justifiably be used to ground empirical work and practice in convivial AI. For example, Actor-Network Theory (ANT), Social Practice Theory (SPT) and Self-determination Theory (SDT) all provide insights on relationality and autonomy that could frame "the triadic relationship between persons, tools, and a new collectivity"[9] in Illich's terms, and underpin convivial tool design. Although these theories don't centre human communities, relationships and power balance in precisely the way Illich does, they nevertheless allow for a consideration of relationships, larger human networks, tools, and/or the meanings that emerge through practices that involve both people and tools. Although a full exploration of these is out of scope, we elaborate on SDT below as it also provides a series of measures that can serve work on convivial CAI.

SDT is a theory of motivation and wellbeing that has at its core, three fundamental psychological needs, two of which are also core to conviviality: autonomy and relatedness. To this, SDT literature adds the construct of 'competence' which, although not an explicit focus of conviviality, is necessary to effective tools use. Over the past few decades, work on SDT (e.g. within education, health and business) has demonstrated that wellbeing is contingent on the satisfaction of these three basic psychological needs (BPNs) and [22] their essential and mediating role in wellbeing has been empirically demonstrated across domains, cultures and age groups. More recently this research has received attention within technology design (see [19] for a review) as a an evidence-based way to design for wellbeing. In brief, within SDT, *autonomy* is conceptualised as willingness, *competence* as feeling capable and effective and *relatedness* as feeling meaningful connection to others. It's worth noting that Autonomy in SDT does not imply individualism. Indeed, it is defined as willingness rather than as control or independence. As such, an individual can very willingly, i.e. *autonomously,* take action for collective benefit, relinquish control, or prioritise others.

Illich instructs us to: "give people tools that guarantee their right to work with high, independent efficiency, thus simultaneously eliminating the need for either slaves or masters and enhancing each person's range of freedom." [9, Ch 2]. This instruction can be interpreted in SDT terms as preserving both the autonomy and competence of technology users.

SDT studies explore the phenomenological experience of how and when environments support or hinder the satisfaction of these needs and in doing so, has developed a series of validated instruments for the measurement of BPN satisfaction and frustration. In conceptualising both autonomy and relatedness robustly, providing evidence for their centrality in flourishing, and providing validated instruments for their measure, SDT provides an evidence-based pathway to operationalising Illich's notion of convivial tools. However, there are some limits to the bridge it can provide. While BPNs have proven valuable in measuring the impact of technology interactions on psychological wellbeing, existing measures rely on individual self-report and don't, therefore, assist with group-level analysis.  Work on SDT within HCI [18, 20] has taken steps to incorporate broader systems of impact by articulating various "Spheres of Technology Experience" including society. However, the design of convivial tools, and indeed convivial conversational agents, will need methods that aim to understand and improve *relationships* (e.g. a dyad, family or community) which points to the group as a unit of analysis—a focus not often taken in technology design.  We might even entertain the notion that convivial design could require a new literature in Relationship-Computer Interaction or even Community-Computer Interaction to enrich existing work in HCI.

In more concrete terms, we can consider the example of a conversational agent discussing reproductive rights with an adolescent. How should it frame the conversation or make recommendations? Without a sense of the family circumstances, existing relationships and the various values that make up the adolescents' environment, it would be easy for the CAI to make suggestions that were insensitive or harmful. Family values, parental ability to support a young person's decisions (emotionally and materially), the consequences any decisions will have on other members of the family, are all interdependent factors. Health decisions are rarely individual decisions, and more often, are highly relational.

Similarly, and coming back to dementia care, a CAI designed to discuss living arrangement options in old age is intervening in the restructuring of relationships. Living arrangements are wrapped up in cultural expectations, values, personality dynamics, and logistical realities that all shape decision-making. A conversational agent would need to be attuned to these variables and the broader network of impact or risk having a negative effect on families and communities.

As such, in order to capture autonomy and relatedness at both an individual and relationship level, SDT measures will likely need to be complemented by qualitative frameworks that address the group level. For example, Delles and Mert propose a technology assessment framework based on conviviality for the context of introducing water technologies into communities. Future work in HCI might draw on parallel work in anthropology, sociology and STS to develop analogous frameworks for convivial technologies in areas such as health.

## Existing Examples of Convivial Tools

Any move toward new paradigms in technology benefits from examples. There are a handful of experimental CAI systems for health reported in the HCI literature that, although not explicitly convivial, could be identified as early movers in the convivial CAI direction. These are typically projects that describe systems intentionally designed for (and evaluated on) the quality of human-to-human relationships. They often centre dyads or groups (rather than individual users) and involve participation from stakeholders throughout the process.

As one example in CAI for health, Seah [24] reported on a system designed to have a positive impact on the relationships between people living with dementia (PLWD) and their carers. The authors developed a set of voice-assistant-based interventions (i.e. 'Alexa skills') to support the mindfulness practices of dyads (people with early-stage dementia and caregivers). Experts in dementia and mindfulness, as well as carers and patients, all contributed to the iterative design and evaluation of the conversational agent. In this example, relationships were centred and participatory methods helped preserve autonomy, thus supporting these two pillars of conviviality.

A second example comes, perhaps surprisingly, from the modern social media environment, and provides a number of insights for convivial design, in particular, for the *coexistence* component. The Change My View (CMV) community on the popular Reddit conversational platform [2] provides a supportive environment for the discussion of contentious topics and opposing views in a way that supports autonomous participation while preserving respectful tolerance. It presents a rare example to counter the divisiveness often amplified by AI-driven fake news engines and recommendation systems. Relatedly, a recent study (consequently criticised for its unethical practices) demonstrated how the thoughtless introduction of CAI could undermine conviviality within such a community [16]. The study researchers created fake LLM-driven accounts (without consent) that joined the community pretending to be people from minoritised groups and even claiming to have lived experience of trauma. This unethical deception points to ways CAI can damage trust among humans within a human community (as we lose the ability to assume genuine humanness). A convivial lens, with its emphasis on community, relationships and individual autonomy would surely have stopped such an application of CAI long before deployment.

In a final techno-moral twist, the LLM used to "role play" vulnerable human beings in the Reddit community was itself trained on text from that same community; Reddit previously licensed its content to OpenAI who used CMV data to train its models. Untangling autonomy from within such a complex web of power, incentives and actors is far from easy. Business models are often in opposition to best practices for caring, coexistence and human autonomy, since generosity, restraint, and mindful attention are seldom supported over compulsive engagement in a profit-driven context. This makes a convivial lens on conversational agent design and deployment all the more important.

## Conclusion: toward convivial design and engineering

While design confronts socio-technical tensions to improve the status quo, engineering seeks ways to leverage technology to scale up services and meet needs. Both disciplines will need to come together for a research program with the ambition to understand the qualities of convivial conversational systems. Based on the existing research as described above, as well as on our own experience building conversational agents for health, we provide four key recommendations for the design of convivial conversational agents below.

**Recommendation 1: Understand both individual and group perspectives and dynamics**
Voinea [28] argued for a systems approach to understanding society, a way to combine the individualist perspective of psychology (and much of technology design), with a holistic view that focuses on the group at the cost of the individual. We recommend the use of SDT methods to measure the autonomy and relatedness components of convivial experience at an individual level. However, new approaches may be required to enrich this with insights only available by taking a collective community lens. Literature in sociology, anthropology and Indigenous methodologies are likely to be of great value to HCI in developing such approaches.

**Recommendation 2: Assess the impact of automation on human relationships**
Understanding how automation may rewire human relationships requires in-depth understanding of the nuances, dynamics and critical benefits of those relationships. As such, future work on supporting conviviality might work with dyads or groups, rather than individuals, for example, in interviews, co-design workshops or via cultural probes. Moreover, ethnography can be valuable to observe relational dynamics in naturalistic settings and may unearth important tacit benefits of care work that, if not identified, could get lost in the process of automation.

**Recommendation 3: Include stakeholders throughout design**
In the health context, participatory and iterative design can be particularly challenging. Patients and their families are often difficult to reach, time poor and/or have access constraints. Moreover, flexible participatory approaches often go against the medical tradition of lengthy and fixed controlled trials with strict pre-defined quantitative outcomes. While patient autonomy has been given more attention within healthcare, relationality, including an understanding of how carers, families and communities are interdependent when it comes to health, gets much less attention. Despite these obstacles, it's hard to imagine how a convivial AI tool could ever be successful without participation from the people whose relationships, values, goals and willingness must be nourished by it. As such we must continue to seek creative ways to partner with stakeholders within health contexts and to approach such relationships in the same way we imagine our technologies should: with support for coexistence, autonomy and relatedness.

**Recommendation 4: Reframe the problem**
Because of the entrenched assumptions that drive technology research and development, convivial tool development will require a mental pivot. As technologists, we need to move from the position of: "how can we scale up by automating human tasks" to "how can automation support humans to have wider reach". A genuine reframing will allow for the possibility that technology may not be called for, or even that *increasing* the number of humans may be the best solution. For example, in a growing number of countries, the best solution to scaling up public health has been the introduction of community health workers – humans, not agents. The community health worker movement, perhaps most famously successful in India, but also in Brazil, Peru and elsewhere, has helped to fill a care gap, not by deploying human replacements but by creating new human roles that require less expertise. These community-based caring roles have allowed, not only tens of thousands of citizens to access care, but also thousands of others to enhance community cohesion by taking on roles that improve public health, and connect young and old throughout their neighbourhoods. It's important to note that this approach does benefit from technology [10, 12], but critically, technology isn't used to replace people, but to innovatively increase their capacity, efficacy and reach.

We hope these recommendations, which recast Illich's notion of Convivial Tools into the modern generative AI landscape, can invigorate new research on conversational technologies that are designed, first and foremost, around human autonomy, human relationships and community.

# Acknowledgements

RAC is funded by: National Institute for Health and Care Research, U.K.: NIHR150287 IMPACT: Innovations using Mhealth for People with dementiA and Co-morbidiTies, the UK Dementia Research Institute, Care Research & Technology Centre. RAC and DP are funded by The Leverhulme Centre for the Future of Intelligence.

# References

[1] Bennett D, Oussama Metatla, Anne Roudaut, and Elisa D. Mekler. 2023. How does HCI Understand Human Agency and Autonomy? In *Proceedings of the 2023 CHI Conference on Human Factors in Computing Systems*, April 19, 2023. ACM, Hamburg Germany, 1–18. https://doi.org/10.1145/3544548.3580651

[2] Burton JW, Ezequiel Lopez-Lopez, Shahar Hechtlinger, Zoe Rahwan, Samuel Aeschbach, Michiel A Bakker, Joshua A Becker, Aleks Berditchevskaia, Julian Berger, Levin Brinkmann, and others. 2024. How large language models can reshape collective intelligence. *Nature human behaviour* 8, 9 (2024), 1643–1655.

[3] Cipolla C. 2009. Relational services and conviviality. *Designing services with innovative methods* 1, (2009), 232–243.

[4] Darzi A. 2024. *Independent investigation of the national health service in England*. Department of Health and Social Care.

[5] Escobar A. 2018. *Designs for the pluriverse: Radical interdependence, autonomy, and the making of worlds*. Duke University Press.

[6] Goodyear P. 2011. Affect, technology and convivial learning environments. In *New perspectives on affect and learning technologies*. Springer, 243–254.

[7] Guo Z, A. Lai, J.H. Thygesen, J. Farrington, T. Keen, and K. Li. 2024. Large Language Models for Mental Health Applications: Systematic Review. *JMIR Mental Health* 11, (2024). https://doi.org/10.2196/57400

[8] Hollingworth S and Ayo Mansaray. 2012. Conviviality under the cosmopolitan canopy? Social mixing and friendships in an urban secondary school. *Sociological research online* 17, 3 (2012), 195–206.

[9] Illich I. 1973. *Tools for conviviality* (First edition ed.). Harper & Row, New York.

[10] Jain S. 2023. How Indian health-care workers use WhatsApp to save pregnant women. *MIT Technology Review*. Retrieved February 20, 2025 from https://www.technologyreview.com/2023/02/03/1067726/indian-healthcare-workers-ashas-whatsapp-misinformation-pregnancy/

[11] Karusala N and Richard Anderson. 2022. Towards Conviviality in Navigating Health Information on Social Media. In *Proceedings of the 2022 CHI Conference on Human Factors in Computing Systems* (*CHI '22*), April 28, 2022. Association for Computing Machinery, New York, NY, USA, 1–14. https://doi.org/10.1145/3491102.3517622

[12] Lúcar MG, José Carlos Vera Tudela, Cecilia Anza-Ramirez, J. Jaime Miranda, and Christopher R. Butler. 2023. Peru initiates the IMPACT project. *The Lancet Neurology* 22, 8 (August 2023), 653. https://doi.org/10.1016/S1474-4422(23)00239-9

[13] Merrilees JJ, Alissa Bernstein, Sarah Dulaney, Julia Heunis, Reilly Walker, Esther Rah, Jeff Choi, Katherine Gawlas, Savannah Carroll, Paulina Ong, and others. 2020. The Care Ecosystem: promoting self-efficacy among dementia family caregivers. *Dementia* 19, 6 (2020), 1955–1973.

[14] Michaelson MG. 1973. Machine-made man. *The New York Times Book Review*, 26–29.

[15] NHS England. Standardising community health services - Planning Guidance. *NHS England*. Retrieved June 4, 2025 from https://www.england.nhs.uk/long-read/standardising-community-health-services/

[16] O'Grady C. 2025. 'Unethical' AI research on Reddit under fire: Ethics experts raise concerns over consent, study design. *Science* (April 2025). https://doi.org/10.1126/science.zzowbgw

[17] Pariser E. 2011. *The filter bubble: What the Internet is hiding from you*. penguin UK.

[18] Peters D, Rafael Alejandro Calvo, and R.M. Ryan. 2018. Designing for motivation, engagement and wellbeing in digital experience. *Frontiers in Psychology* 9, MAY (2018). https://doi.org/10.3389/fpsyg.2018.00797

[19] Peters D and Rafael A. Calvo. 2023. Self-Determination Theory and Technology Design. In *The Oxford Handbook of Self-Determination Theory*, Richard M. Ryan (ed.). Oxford University Press, 0. https://doi.org/10.1093/oxfordhb/9780197600047.013.49

[20] Peters D, Karina Vold, Diana Robinson, Rafael A Calvo, and Senior Member. 2020. Responsible AI-Two Frameworks for Ethical Design Practice. *IEEE TRANSACTIONS ON TECHNOLOGY AND SOCIETY* 1, 1 (2020). https://doi.org/10.1109/TTS.2020.2974991

[21] Ryan RM and Edward L. Deci. 2017. Human autonomy: Philosophical perspectives and the phenomenology of self. In *Self-Determination Theory: Basic psychological needs in motivation, development, and wellness*. 51–79.

[22] Ryan RM and Edward L Deci. 2017. *Self-determination theory: Basic psychological needs in motivation, development, and wellness*. Guilford Publications.

[23] Sadek M, Rafael A. Calvo, and Céline Mougenot. 2023. Co-designing conversational agents: A comprehensive review and recommendations for best practices. *Design Studies* 89, (December 2023), 101230. https://doi.org/10.1016/j.destud.2023.101230

[24] Seah C, Sijin Sun, Zheyuan Zhang, Talya Porat, Andrew Waterhouse, and Rafael A Calvo. 2022. Using a User Centered Design Approach to Design Mindfulness Conversational Agent for Persons with Dementia and their Caregivers. In *Adjunct Proceedings of the 2022 ACM International Joint Conference on Pervasive and Ubiquitous Computing and the 2022 ACM International Symposium on Wearable Computers*, 2022. 207–210.

[25] Szondy MB and Ágnes Magyary. 2025. Artificial Intelligence (AI) in the Family System: Possible Positive and Detrimental Effects on Parenting, Communication and Family Dynamics. *European Journal of Mental Health* 20, (2025), 1–8.

[26] Thirunavukarasu AJ, Darren Shu Jeng Ting, Kabilan Elangovan, Laura Gutierrez, Ting Fang Tan, and Daniel Shu Wei Ting. 2023. Large language models in medicine. *Nature medicine* 29, 8 (2023), 1930–1940.
[27] Tiersen F, Philippa Batey, Matthew JC Harrison, Lenny Naar, Alina-Irina Serban, Sarah JC Daniels, Rafael A Calvo, and others. 2021. Smart home sensing and monitoring in households with dementia: user-centered design approach. *JMIR aging* 4, 3 (2021), e27047.
[28] Voinea C. 2018. Designing for conviviality. *Technology in Society* 52, (2018), 70–78.
[29] Weidinger L, Jonathan Uesato, Maribeth Rauh, Conor Griffin, Po-Sen Huang, John Mellor, Amelia Glaese, Myra Cheng, Borja Balle, Atoosa Kasirzadeh, and others. 2022. Taxonomy of risks posed by language models. In *Proceedings of the 2022 ACM conference on fairness, accountability, and transparency*, 2022. 214–229.

## About the Authors

Rafael A. Calvo., PhD is Professor at the Dyson School of Design Engineering, Imperial College London. He is also co-lead at the Leverhulme Centre for the Future of Intelligence (Imperial spoke). He focuses on the design of systems that support wellbeing in areas of mental health, medicine and education, and on the ethical challenges raised by new technologies.

Dorian Peters, PhD is Assistant Professor at the Leverhulme Centre for the Future of Intelligence, University of Cambridge.